\documentclass[aps,pre,10pt,preprint,floatfix]{revtex4-2}
\usepackage{amsmath}
\usepackage{amssymb}
\usepackage{amsfonts}
\usepackage{bm}
\usepackage{graphicx}
\usepackage{color}
\usepackage{lipsum}
\usepackage{lineno}
\usepackage{fullpage}
\usepackage{jabbrv}

\newcommand{\bu}{\boldsymbol{u}}
\newcommand{\eps}{\epsilon}

\newcommand{\dimf}{V}
\newcommand{\f}{f}
\newcommand{\W}{\mathcal{W}}
\newcommand{\M}{\mathcal{M}}

\newcommand{\C}{\mathcal{C}}

\newcommand{\LL}{\mathcal{L}}
\newcommand{\D}{\mathcal{D}}
\newcommand{\Rey}{\text{Re}}
\newcommand{\tr}[1]{{\tilde{{#1}}}}
\newcommand{\dd}{\mathrm{d}}
\newcommand{\Hess}{\mathcal{H}}

\newcommand{\bx}{\boldsymbol{x}}
\newcommand{\ba}{\boldsymbol{\alpha}}
\newcommand{\grad}{\boldsymbol{\nabla}}

\newcommand{\bQ}{\boldsymbol{Q}}
\newcommand{\bC}{\boldsymbol{C}}
\newcommand{\bsL}{\boldsymbol{L}}

\graphicspath{{Figures/}}
\begin{document}
\title{Brachistochronous motion of a flat plate parallel to its surface immersed in a fluid}
\author{Shreyas Mandre}
\email[Corresponding author:]{ shreyas.mandre@warwick.ac.uk}
\affiliation{Mathematics Institute, University of Warwick, Coventry UK CV4 7AL}
\date{\today}

\begin{abstract}
We determine the globally minimum time $T$ needed to translate a thin submerged flat plate a given distance parallel to its surface within a work budget.
The Reynolds number for the flow is assumed to be large so that the drag on the plate arises from skin friction in a thin viscous boundary layer.
The minimum is determined using a steepest descent, where an adjoint formulation is used to compute the gradients.
Because the equations governing fluid mechanics for this problem are nonlinear, multiple local minima could exist.
Exploiting the quadratic nature of the objective function and the constraining differential equations, we derive and apply a ``spectral condition'' to show that the converged local optimum to be a global one.
The condition states that the optimum is global if the Hessian of the Lagrangian in the state variables constructed using the converged adjoint field is positive semi-definite at every instance.
The globally optimum kinematics of the plate starts from rest with speed $\propto t^{1/4}$ and comes to rest with speed $\propto (T-t)^{1/4}$ as a function of time $t$.  
For distances much longer than the plate, the work-minimizing kinematics consists of an optimum startup, a constant-speed cruising, and an optimum stopping.
The spectral condition can also be used for optimization problems constrained by other quadratic partial differential equations.
\end{abstract}
\maketitle

Optimization over flow fields that satisfy equations governing fluid motion has many applications including animal propulsion\cite{Berman2007,Pesavento2009,Tam2011,Alben2013,Kern2006,Vincent2020}, engineering\cite{Ibrahim2001,Terashima2001,Young2014,Economon2015,Eggl2018,Pasche2019}, cardiovascular\cite{He1994,Marsden2014} and sports biomechanics\cite{Labbe2019},  and fundamental fluid mechanics\cite{Pringle2010,Pringle2012,Passaggia2013,Kerswell2014,Eggl2018,Souza2020}.
The solution of the time-dependent nonlinear Navier-Stokes equations coupled with optimization makes these problems formidable. 
Powerful optimization techniques are available in the limit where viscous forces dominate over inertial ones \cite[e.g., see][]{Tam2007,Tam2011}, which renders the governing equations linear and kinematically reversible.
These techniques do not apply to the situation we consider in this article when inertia dominates and the nonlinearities cannot be ignored.
Gradient-descent methods based on calculus of variations are formulated for numerical optimization \cite[e.g.][]{Pringle2010,Pringle2012,Eggl2018,Boujo2019,Pasche2019} where an adjoint formulation is used to calculate the gradient.
However, multiple local optima might exist and the one found by these methods cannot be guaranteed to be global.
Because an exhaustive search in the infinite-dimensional state space is impossible, whether these methods find the global optimum remains unknown.

\begin{figure}[t]
\centerline{\includegraphics[height=20mm]{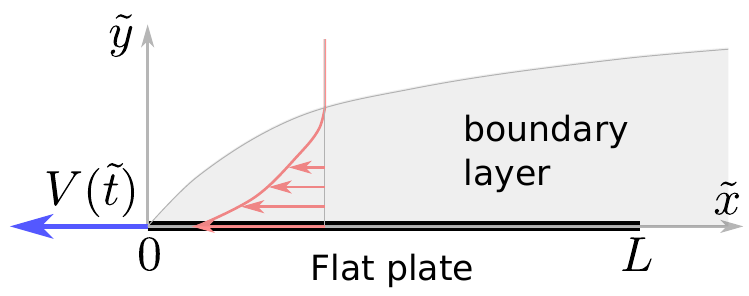}}
\caption{Schematic a flat plate moving through a fluid.}
\label{fig:Schematic}
\end{figure}

Here we solve an instance of the fluid mechanical bracistochrone, a problem that epitomizes the general fluid mechanical optimization by sharing its nonlinear structure.
The problem asks the shortest time to move an object in a fluid a fixed distance within a limited work budget, which we solve for a flat plate moving parallel to its surface.
The problem is equivalent to determining the profile of transient speed $\dimf(\tr{t})$ with time $\tr{t}$ of a flat plate of length $L$ moving parallel to its surface a distance $D$ in time $T$ (see Figure \ref{fig:Schematic}) that minimizes the mechanical work $\tr{\W}$. 
The surrounding fluid of density $\rho$ and viscosity $\mu$ (with $\nu = \mu/\rho$) is of infinite extent and initially static, and the flow is considered laminar.
The plate is infinitesimal in its thickness and infinitely long in the third dimension so that the resulting flow is two-dimensional and confined to a thin layer close to the plate.

The dimensionless parameters $\Rey = \rho DL/(\mu T)$ and $\eps = D/L$ represent the Reynolds number and the target distance to be travelled relative to the plate length.
We consider the regime for $\Rey \gg 1$ and arbitrary $\eps$ thereby retaining the essential nonlinearity in the governing equations.
We find that for $\eps \gg 1$, the minimum work $\tr\W_\text{min}$ needed for travel in fixed duration $T$, or equivalently the shortest time $T_\text{min}$ under a work-budget $\tr\W$, asymptotically approach
\begin{linenomath}
\begin{align}
\label{eqn:mainresult}
\tr{\W}_\text{min} \approx 1.328 \sqrt{\dfrac{\mu \rho D^5L}{T^3}}, \quad
T_\text{min} \approx 1.21 \left(\dfrac{\mu \rho D^5 L}{\tr{\W}^2}\right)^{1/3},
\end{align}
\end{linenomath} 
The details of the adjoint-based gradient descent and characterization of the optimum kinematics are presented in \S\ref{sec:brachistochrone}.

We overcome the primary drawback of gradient descent by deriving a sufficient condition to test whether a local optimizer obtained using the adjoint forumation is a global one.
This is achieved using the Lagrange dual formulation to derive a bound on the objective function that agrees with the calculated optimum.
We start in \S\ref{sec:cfgo} with the derivation of the spectral condition for optimization of quadratic objectives constrained by quadratic partial differential equations, 
The condition is then used in \S\ref{sec:brachistochrone} to prove that the optimum found for the brachistochrone problem is global.

\begin{figure}[t]
\includegraphics[width=3in]{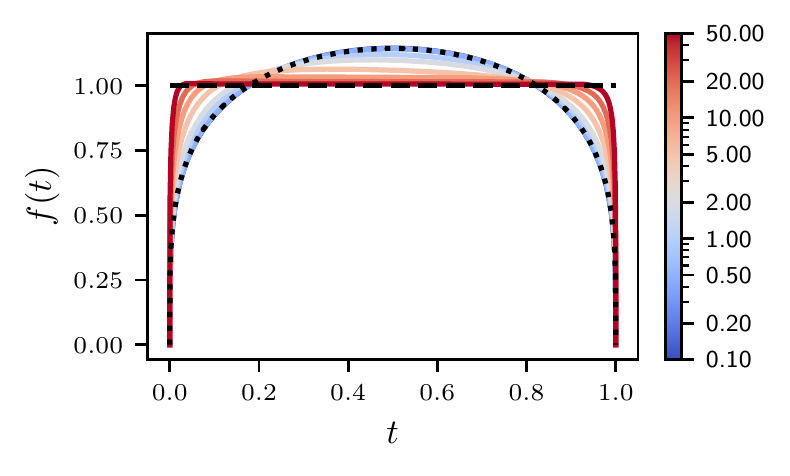}
\caption{Optimal $\f(t)$ for different $\eps$ coded according to the color in the adjoining colorbar. Dotted curve shows $\f_0(t)$ from \eqref{eqn:smallepsf} and the dashed curve shows unity.}
\label{fig:Profiles}
\end{figure}

\section{Spectral condition for global optimality}\label{sec:cfgo}

Consider a time-independent region $\Omega$ occupied by a fluid for a time $t\in[0,T]$.
The objective is to find 
\begin{linenomath}
\begin{subequations} \label{eqn:genopt}
\begin{align}
\W_\text{min} = \min_{\bu(\bx,t)} \W[\bu] \equiv \int_0^T \int_\Omega (q(\bu)+l(\bu))~\dd \Omega~\dd t \label{eqn:genobj} \\
\text{subject to} \quad
 \bu_t + \bQ(\bu) + \bsL(\bu) + \bC = 0, 
 \label{eqn:governing}
\end{align}
\end{subequations}
\end{linenomath}
for $\bx \in \Omega$ and $t\in(0,T)$, where $q(\bu)$ and $\bQ(\bu)$ are quadratic, $l(\bu)$ and $\bsL(\bu)$ are linear, and $\bC(\bu)$ is constant in $\bu$ and its spatial derivatives.
In particular, assume that $q(\bu)$ and $\bQ(\bu)$ do not depent on $\bu_t$.
The optimization is over the initial condition \cite[e.g., finding the minimal seed for transition to turbulence,][]{Pringle2010,Pringle2012}, the boundary condition \cite[e.g., kinematic optimization][]{Eggl2018,Mandre2020} for $\bu$, the forcing $\bC$, or any combination of the above.

The optimization proceeds by writing the Lagrangian 
\begin{linenomath}
\begin{align}
\LL[\bu, \ba] = \int_0^T \int_\Omega \left[ q + l - \ba\cdot (\bu_t + \bQ + \bsL + \bC) \right]~\dd \Omega~\dd t,
\end{align}
\end{linenomath}
where $\ba(\bx, t)$ is the adjoint field.
Accounts of how to find the adjoint, the gradients $\delta \LL/\delta \bu$, and descend along it, can be found elsewhere\cite[e.g.,][]{He1994,Mohammadi2004,Pringle2010, Pringle2012,Passaggia2013}. 
Suppose this process converges to $\bu=\bu_*$, corresponding to $\ba = \ba_*$ and $\W = \W_*$. 

To test if the converged state is a global minimum, we construct the Lagrange dual $\D[\ba] = \inf_{\bu} \LL[\bu, \ba]$.
The value of $\D[\ba]$ for any $\ba$ is a lower bound on $\W[\bu]$ for $\bu$ that satisfies \eqref{eqn:governing} \cite[e.g., see Ref.][for a proof]{Boyd2004}.
We choose $\ba = \ba_*$. 
To determine $\D[\ba_*]$, we need to minimize $\LL[\bu, \ba_*]$ over all $\bu$.
Since both $\W$ and \eqref{eqn:governing} are quadratic in $\bu$, the first variation conditions are linear in $\bu$, which are satisfied by $\bu_*$.
Therefore, the second variation 
\begin{linenomath}
\begin{align}
\Hess[\bu, \ba_*] = \int_0^T \int_\Omega (q - \ba_*\cdot \bQ)\dd\Omega\dd t
\end{align}
\end{linenomath}
completely determines whether $\bu_*$ is the minimizer of $\LL[\bu, \ba_*]$.
This leads to
\begin{linenomath}
\begin{align}
\D[\ba_*] = 
\begin{cases}
\W_* &\text{ if } \Hess[\bu, \ba_*] \geq 0 \text{ for all } \bu, \\
-\infty &\text{ otherwise.}
\end{cases}
\end{align}
\end{linenomath}
Since $\bu_t$ does not appear in the integrand for $\Hess$, the second variation is positive semi-definite if and only if 
\begin{linenomath}
\begin{align}\label{eqn:genspectral} 
 \int_\Omega (q - \ba_*\cdot \bQ)~\dd\Omega \geq 0 \text{ for all } \bu \text{ and every } t.
\end{align}
\end{linenomath}
If \eqref{eqn:genspectral} is satisfied then $\W_*$ is a lower bound on $\W[\bu]$ which is attained at $\bu_*$ and, therefore, the global minimum. 
The objective function can also contain terms arising from the integral of quadratic functions of $\bu$ and its spatial derivatives over the boundaries of $\Omega$, and/or at the initial and final time, which we have not explicitly written here.
If so, these may also contribute to the Hessian.

As an example, if \eqref{eqn:governing} represents the incompressible Navier-Stokes equations, then $\bu$ is the divergence-free velocity field, $\bQ=\bu \cdot\grad\bu + \grad p $ represents the advection term, $\bsL = -\nu \nabla^2 \bu$ represents the viscous term and $\bC$ the body force term.
For objectives such as the dissipation rate, $q(\bu) = |\grad\bu|^2$ and $l(\bu)=0$, the assumptions that the nonlinearities be quadratic and independent of $\bu_t$ are satisfied.
The corresponding condition for global optimality is 
\begin{linenomath}
\begin{align}\label{eqn:spectral}
 \int_\Omega \left( |\grad\bu|^2 - \ba_* \cdot (\bu\cdot\grad) \bu \right)~\dd\Omega \geq 0,
\end{align}
\end{linenomath}
over all divergence-free $\bu$, applied separately at each $t$.
A condition equivalent to \eqref{eqn:spectral} was named as the ``spectral constraint'' \cite{Doering1994,Kerswell1999} because it requires that the self-adjoint linear operator underlying the quadratic form, i.e. the Hessian of $\LL[\bu,\ba_*]$, have non-negative eigenvalues.
Based on this nomenclature, we term \eqref{eqn:genspectral} the spectral condition.

Failure to satisfy the condition does not necessarily mean that $\bu_*$ is not global but could imply a duality gap\cite{Boyd2004}.
A duality gap exists when $\D[\ba] < \W_\text{min}$ for all $\ba$.
Conversely, satisfaction of the spectral condition demonstrates the absence of a finite duality gap.

\begin{figure}[t]
\includegraphics{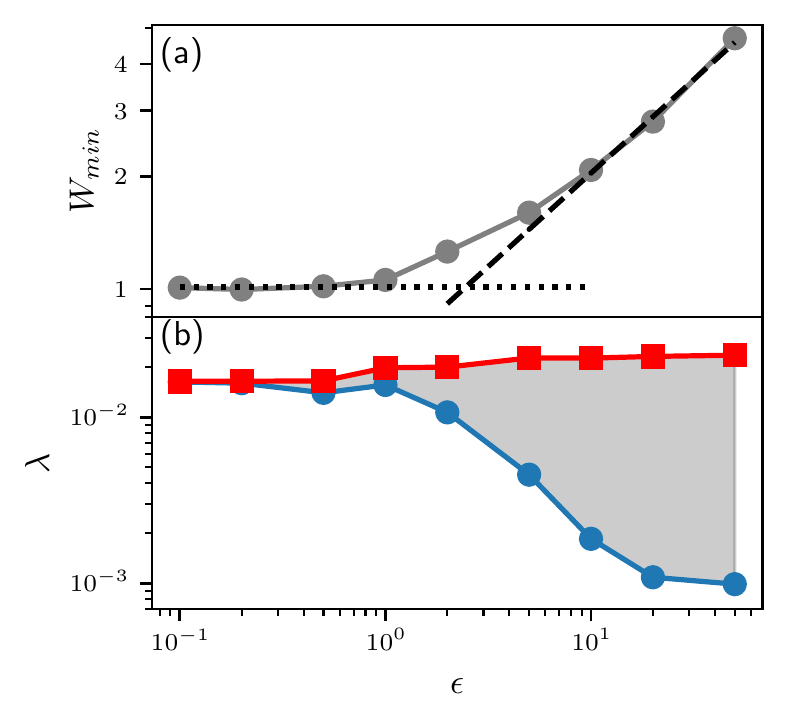}
\caption{Minimum work for translating a flat plate and the verification of the corresponding spectral constraint. (a) $\W_\text{min}$ as a function of $\eps$. The dotted line shows $\W_\text{0,min}\approx 1.014$ from \eqref{eqn:smallepsW} and the dashed lines hows $0.664 \eps^{1/2}$. (b) Minimum $\lambda_\text{min} = \min_{t\in(0,1)} \lambda(t)$ (blue circles) and maximum $\lambda_\text{max} = \max_{t\in(0,1)} \lambda(t)$ (red squares) of $\lambda(t)$. 
}
\label{fig:Plots}
\end{figure}

\section{Work-minimizing kinematics of a flat-plate} \label{sec:brachistochrone}

\subsection{Mathematical formulation}\label{sec:mathform}
In the limit $\Rey \gg 1$, the problem is formulated based on a thin viscous boundary layer that governs the drag.
The fluid outside this layer, to leading order in $\Rey$, remains stationary as the plate moves.
To model the flow in boundary layer, we use a coordinate system attached to the leading edge of the plate, as shown in Figure~\ref{fig:Schematic}, and a reference frame attached to farfield stationary fluid.
Exploiting the reflection symmetry, we only consider the flow field for $y\geq 0$ and the drag on one face of the plate.
The coordinates $\tr{x}$, $\tr{y}$ and the flow velocity $(\tr{u}(\tr{x},\tr{y},\tr{t}), \tr{v}(\tr{x},\tr{y},\tr{t}))$ in the boundary layer are non-dimensionalized as
\begin{linenomath}
\begin{align}
 \tr{t} = Tt,~\tr{x} = Lx,~ \tr{y} = \delta y,~\tr{u} = \dfrac{D}{T} u,~\tr{v} = \dfrac{D\delta}{TL} v,
\end{align}
\end{linenomath}
where $\delta = \sqrt{\nu T}$ to yield the dimensionless form of the governing boundary layer equations as
\begin{linenomath}
\begin{subequations}
 \label{eqn:prandtl}
\begin{align}
 \M &\equiv u_t + \eps (u+\f)u_x + \eps v u_y - u_{yy} = 0, \\ 
 \C &\equiv u_x + v_y = 0, \label{eqn:prandtlmass}
\end{align}
\end{subequations}
\end{linenomath}
in the semi-infinite half-space $y\geq 0$ and
\begin{linenomath}
\begin{subequations}
\label{eqn:bcs}
\begin{align}
 u|_{x\in P, y=0} + \f(t) = u_y|_{x \not\in P, y=0} = 0, \label{eqn:plate}\\
 v|_{y=0} = u_y|_{y=\infty} = u|_{x=\pm\infty} = u|_{t=0} = 0,\label{eqn:init}
\end{align}
\end{subequations}
\end{linenomath}
where $P = [0,1]$ specifies the extent of the flat plate, $\f(t) = (T/D)\dimf(\tr{t})$ is the dimensionless plate speed.
Subscripts $x$, $y$ and $t$ on $u$ and $v$ denote partial derivatives.

The following notation will be also used
\begin{linenomath}
\begin{align}
 \langle \phi \rangle_{\gamma_1 \dots \gamma_m} = \iiint \phi~\dd \gamma_1 \dots \dd \gamma_m \dd t,
\end{align}
\end{linenomath}
where $\gamma_i$ could be any of $x$, $y$ or $t$. 
The limits of integration are $[0,1]$ on $t$, $[0,\infty]$ on $y$ and $[-\infty, \infty]$ on $x$. 
For example, $\langle \phi \rangle_{xy} = \int_0^\infty \int_{-\infty}^\infty \phi~\dd x \dd y$. 
For simplicity, $\langle \phi \rangle = \langle \phi \rangle_{xyt}$. 

The work done is $\tr{\W} =  D^2 L \sqrt{{\mu \rho}/{T^3}}~\W[f]$, where $\W[\f] = \langle \f(t) u_y(x,y=0,t)\rangle>_{xt}$.
For $(u,v)$ satisfying (\ref{eqn:prandtl}-\ref{eqn:bcs}), the work done must appear as an increase in the kinetic energy of the fluid or be viscously dissipate, i.e., 
\begin{linenomath}
\begin{align}
 \W[\f] = \hat{\W}[\f] \equiv \left\langle \tfrac{1}{2} \left. u^2\right|_{t=1} \right\rangle_{xy} + \langle u_y^2 \rangle
 \label{eqn:energy}
\end{align}
\end{linenomath}

The objective of the optimization is to minimize $\W[f]$ subject to $\M = \C= 0$ in the fluid.
The Lagrangians using $\W[f]$ and $\hat{\W}[\f]$ are
\begin{linenomath}
\begin{subequations}
\label{eqn:Lagrangian}
\begin{align}
\LL &= \W[\f] - \langle \alpha \M + \eps \beta \C \rangle - \lambda \D, \quad \text{and} \label{eqn:Lag} \\
\hat{\LL} &= \hat{\W}[\f] - \langle \hat\alpha \M + \eps \beta \C \rangle - \lambda \D, \label{eqn:Laghat}
\end{align}
\end{subequations}
\end{linenomath}
respectively, where $\alpha(x,y,t)$, $\hat\alpha(x,y,t)$, $\beta(x,y,t)$ and $\lambda$ are Lagrange multipliers, and $\D = \langle f \rangle_t - 1$ is the constraint on the distance travelled by the plate.
We also define $\hat{u} = u + f$, which is the $x$-component of the fluid velocity in the reference frame of the plate.
Gradient descent is convenient using $\LL$ because in these variables the adjoint satisfies an equation similar to the $u$.
On the other hand, the spectral condition is derived using $\hat{\LL}$, which when expressed in $\hat{u}$ reveals an eigenvalue problem with homogenous boundary conditions.
By virtue of \eqref{eqn:energy}, the two formulations in $\LL$ and $\hat{\LL}$ are equivalent, related by $\hat\alpha = \alpha + u$. 

\begin{figure}
\includegraphics{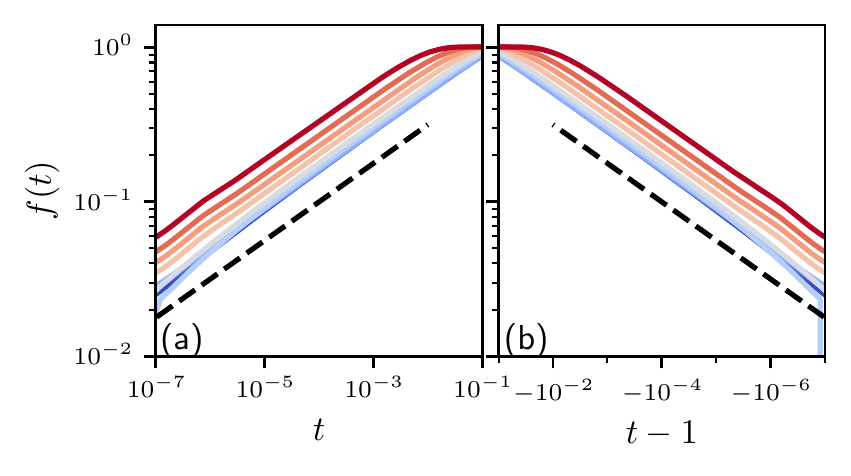}
\caption{Optimal starting and stopping kinematics plotted on a logarithmic scale. (a) Starting kinematics. Color code identical to that in Figure~\ref{fig:Profiles}. Dashed line shows $t^{1/4}$. (b) Stopping kinematics. Dashed line shows $(T-t)^{1/4}$.}
\label{fig:Logscale}
\end{figure}

The multipliers $\alpha$ and $\beta$ satisfy the condition that first variations of $\LL$ due to $u$ and $v$ vanish, i.e.,
\begin{linenomath}
\begin{subequations}
\begin{align}
\alpha_t + \eps (u+f) \alpha_x + \eps (v\alpha)_y + \eps \beta_x + \alpha_{yy} = 0, \label{eqn:alpha} \\
\beta_y = \alpha u_y, \label{eqn:beta}
\end{align}
\label{eqn:adjoints}
\end{subequations}
\end{linenomath}
in $y>0$ subject to the boundary conditions
\begin{linenomath}
\begin{subequations}
\begin{align}
\alpha|_{x \in P, y=0} - \f (t) = \alpha_y|_{x \not\in P, y=0} = 0, \label{eqn:aplate} \\
\beta|_{y=\infty} = \alpha_y|_{y=\infty} = \alpha|_{x=\infty} = \alpha|_{t=1} = 0. \label{eqn:asym}
\end{align}
\label{eqn:abcs}
\end{subequations}
\end{linenomath}
The first variation with respect to $f$ given by
\begin{linenomath}
\begin{align}
\dfrac{\delta \LL}{\delta \f} = \left\langle \left[u_y - \alpha_y\right]_{y=0} \right\rangle_x - \eps \langle \alpha u_x \rangle_{xy} - \lambda
\label{eqn:gradf}
\end{align}
\end{linenomath}
must also vanish.

The optimization is carried out using gradient descent in $\f$ using \eqref{eqn:gradf} while solving \eqref{eqn:prandtl}, \eqref{eqn:bcs}, \eqref{eqn:adjoints} and \eqref{eqn:abcs} numerically as described in Materials and Methods.
We find this procedure converges to $f=f_*$, $u=u_*$, $v=v_*$, $\alpha = \alpha_*$ and $\W[f]=\W_\text{min}$, as $\delta \LL/\delta \f$ approaches 0.
The converged profiles for $f(t)$ are shown in Figures~\ref{fig:Profiles} and the corresponding $\W_\text{min}$ in \ref{fig:Plots}(a).

\subsection{Spectral condition} To prove that the computed minimum is global, we choose $\hat\alpha=\hat\alpha_* = \alpha_* + u_*$ and the Lagrange dual
\begin{linenomath}
\begin{align}\label{eqn:LagDual}
 \D[\hat\alpha] \equiv \min_{u,v,\beta,\f} \hat{\LL}.
\end{align}
\end{linenomath}
Because $\hat{\LL}$ is quadratic in $(u,v,\beta,f)$, by virtue of Eqs. (\ref{eqn:adjoints}-\ref{eqn:abcs}) $(u_*,v_*, \beta_*,f_*)$ is a stationary point of $\hat{\LL}$ in those variables, for which $\hat{\LL} = \W_\text{min}$. 
It is the second variation 
\begin{linenomath}
\begin{align}
 \Hess[\hat\alpha_*; u,v,f] = \W[f] - \eps \left\langle \hat\alpha_* ((u+f)u_x + vu_y) \right\rangle
\end{align}
\end{linenomath}
in $(u,v,f)$ for $(u,v)$ satisfying \eqref{eqn:prandtlmass} that determines whether $(u_*,v_*, \beta_*,f_*)$ is a minimum of $\hat\LL$.
This leads to 
\begin{linenomath}
\begin{align} \label{eqn:DualValue}
\W[f] \geq \D[\hat\alpha_*] = 
 \begin{cases}
  \W_\text{min} &\text{if } \Hess \geq 0 \text{ for all } (u,v,f), \\
  -\infty       &\text{otherwise}.
 \end{cases}
\end{align}
\end{linenomath}
The condition $\Hess \geq 0$ is satisfied if for each $t\in (0,1)$
\begin{linenomath}
\begin{align}
\label{eqn:spectralprandtl}
\varDelta[u,v] = \left\langle \hat u_y^2 -\hat\alpha_* \eps (\hat u \hat u_x + v \hat u_y) \right\rangle_{xy} \geq 0,
\end{align}
\end{linenomath}
for all $(\hat u,v)$ satisfying \eqref{eqn:prandtlmass} and \eqref{eqn:bcs}. 

\eqref{eqn:prandtlmass} implies a stream function $\psi$ such that $\hat{u} = \psi_y$ and $v = -\psi_x$.
Substituting this in \eqref{eqn:spectralprandtl} and integrating by parts yields the equivalent optimization problem
\begin{linenomath}
\begin{align}
\label{eqn:spectraloptimization}
\begin{split}
 \lambda(t) = &\min_\psi \varDelta = \min_\psi \left\langle \psi_{yy}^2 +\eps \hat\alpha_{*x} \psi_y^2 + \eps \hat\alpha_{*y} \psi_x\psi_y \right\rangle_{xy}, \\
 &\text{subject to } \left\langle \psi^2 + \psi_x^2 + \psi_y^2 \right\rangle_{xy} = 1
\end{split}
\end{align}
\end{linenomath}
for each $t\in(0,1)$.
The solution of this problem implies $\varDelta \geq \lambda(t) \left\langle \psi^2 + \psi_x^2 + \psi_y^2 \right\rangle_{xy}$ and hence the spectral constraint is satisfied if $\lambda(t)\geq 0$ for $t\in(0,1)$.
Here $\lambda(t)$ is determined numerically by constructing the generalized eigenvalue problem from the linear operators underying the quadratic forms in \eqref{eqn:spectraloptimization}, which is described in Materials and Methods.
As shown in Figure~\ref{fig:Plots}(b), $\lambda(t)$ is positive for each $t$, thus proving that the local minimum found in \S\ref{sec:brachistochrone}\ref{sec:mathform} is a global one.
\subsection{Interpretation of the results}
As seen in Figures~\ref{fig:Profiles} and \ref{fig:Plots}(a), the analytical solution~\cite{Mandre2020} derived for $\eps \ll 1$,
\begin{linenomath}
\begin{subequations}\label{eqn:smalleps}%
\begin{align}
\f(t) = \f_0(t) = C {t^{1/4} (1-t)^{1/4}} 
\text{ and }  \label{eqn:smallepsf} \\
\W_\text{min} = \W_\text{0,min} \approx 1.014, 
\label{eqn:smallepsW}
\end{align}
\end{subequations}
\end{linenomath}
where $C \approx 1.62$, approximates the solution for finite $\eps\leq 0.5$. 
As $\eps$ increases beyond, the optimal $\f(t)$ departs from $\f_0(t)$ while $\W_\text{min}$ rises above $\W_\text{0,min}$.
In particular, $\f(t)$ starts from $\f(0) = 0$ and end at $\f(1) = 0$ but flattens out in the middle. 
For $\eps\gtrsim 5$, $\f(t)$ approaches unity, except for near the start and the end.
In other words, the optimum kinematics to cover a distance $D\gg L$ in time $T$ is to cruise at the average speed $U\approx D/T$, except to start and stop.
For $\eps \gg 1$, $\W_\text{min}$ is also observed to rise $\propto \eps^{1/2}$.

\begin{figure}
\includegraphics[width=3in]{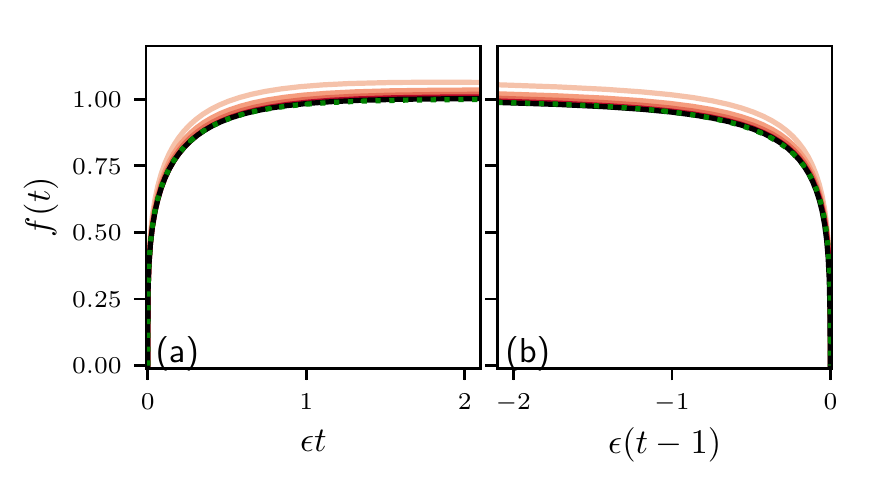}
\caption{Optimal starting and stopping kinematics. (a) The start-up dynamics of optimal kinematics for $\eps$=5, 10, 20 and 50 (same color code as in Figure~\ref{fig:Profiles}) plotted against $\eps t$. Solid black curve shows the optimal starting kinematics and the overlapping dotted green curve the empirical fit in \eqref{eqn:empiricalfit}. (b) Same as (a) but for stopping kinematics.}
\label{fig:Collapse}
\end{figure}

The following argument rationalizes these observations.
The drag according to Blasius\cite{Blasius1907,Batchelor1967} for a flat plate moving at steady speed $U$ is given by $0.664 \sqrt{\mu \rho U^3 L}$, and consequently, the work done to move the plate is $\tr{\W}_\infty = 0.664 T \sqrt{\mu \rho U^5 L}$.
Consider covering the distance $D=UT$ in two stages of duration $aT$ and $(1-a)T$ moving with speeds $U_1 = bU/a$ and $U_2 = (1-b)U/(1-a)$, respectively, for constants $a$ and $b$ between 0 and 1.
When $\eps \gg 1$, steady state is reached much faster than the duration of each segment, and the modified kinematics approximately incurs the work
\begin{linenomath}
\begin{align}
0.664 T \sqrt{\mu \rho U^5 L} \left(\dfrac{b^{5/2}}{a^{3/2}}  + \dfrac{(1-b)^{5/2}}{(1-a)^{3/2}} \right).
\end{align}
\end{linenomath}
This work is minimized when $b=a$, or $U_1 = U_2 = U$.
In other words, the penalty incurred in the work done when traveling fast outweighs the benefits accrued when traveling slower, explaining why the optimum avoids modulation of the speed.
Converting $\tr{\W}_\infty$ to a dimensionless form yields $\W_\text{min} \approx 0.664 \eps^{1/2}$ for $\eps \gg 1$, agreeing up to the leading order with the results of the computations as shown in Figure~\ref{fig:Plots}(a).
Accounting for the two sides of the plate leads to \eqref{eqn:mainresult}.

It is readily seen why the optimal profile avoids an impulsive start and stop.
For $t\ll 1$ (and $1-t\ll 1$) owing to the development of the viscous boundary layer, the unsteady inertia $u_t$ and shear viscosity $u_{yy}$ in \eqref{eqn:prandtl} dominate, while advection $\eps (u+\f)u_x + \eps vu_y$ is negligible.
Therefore, in the first variation condition \eqref{eqn:gradf}, the dominant balance is between $u_y$ and $\alpha_y$.
For an impulsive start, the initial shear stress profile on the plate, $u_y|_{y=0}\approx f(0)/\sqrt{\pi t}$ due to the growth of the boundary layer thickness proportional to $\sqrt{t}$.
The adjoint dynamics, due to their backwards evolution in time, does not ``know'' about the impulsive start, and therefore cannot generate an $\alpha_y$ that matches this asymptotic behavior. 
Therefore, for small $t$, one can always reduce the work done by eliminating the impulsive start (and analogously for an impulsive stop).

We also find that for finite $\eps$ the optimal starting and stopping dynamics behave proportional to $t^{1/4}$ and $(1-t)^{1/4}$, respectively.
This is observed in the numerical solution for over four orders of magnitude in $t$, as shown in Figure~\ref{fig:Logscale}.
The reason is analogous to that in the analytical solution for vanishing $\eps$ in \eqref{eqn:smalleps}.
Near the starting and stopping time, the advection is negligible and the optimal kinematics are governed by the viscous diffusion of momentum within the fluid, which causes this behavior\cite{Mandre2020}.

Finally, we observe that for $\eps \gg 1$, the starting and stopping dynamics as a function of $\eps t$ are independent of $\eps$, as seen in Figure~\ref{fig:Collapse}.
(Here $\eps t$ is the ratio of the distance covered when travelling at speed $D/T$ to the plate length.) 
For $\eps \geq 5$, these optimal starting and stopping dynamics approach successively closer to limiting curves.
These limiting curves denote the work-minimizing kinematics for the flat plate to attain a constant cruising speed from rest and to stop from the cruising motion, respectively.
These starting and stopping kinematics do not depend separately on the total distance travelled $D$ or the time taken $T$, but depend only on the cruising speed.
For the purpose of determining the optimal start-up and stopping kinematics, we non-dimensionalize the target cruising speed to unity.
The governing equations to determine these kinematics are identical to the ones developed in \S\ref{sec:brachistochrone}\ref{sec:mathform} with the following exceptions.
The variables $t$ and $y$ are rescaled as $\eps t = t'$ and $y\sqrt{\eps} = y'$, and formally $t'$ ranges from 0 to $\tau \to \infty$ (all the integrals in Eq. \ref{eqn:Lagrangian} are  now over the semi-infinite time interval).
The distance travelled condition $\D=0$ is replaced by the unit target cruising speed
\begin{linenomath}
\begin{align}
 \D' = \lim_{\tau \to \infty} \dfrac{1}{\tau} \int_0^\tau \f_\text{start}(t') \dd t' - 1 = 0.
\end{align}
\end{linenomath}
An additional constraint that the final velocity profile approaches the Blasius steady-state profile\cite{Batchelor1967} $u_s(x,y,t)$ with $u_s(x\in P, y=0, t) = -1$ is added.
Imposing a target final state for $u$ implies that the condition for starting the backwards time-integration of $\alpha$ must be determined as part of the solution.
This condition is trivially determined to be the steady solution of \eqref{eqn:adjoints} with $u = u_s$, $u(x\in P,y=0,t)=-1$ and $\alpha(x\in P, y=0,t) = 1$.
The numerical procedure described in Materials and Methods then yields the optimum start-up kinematics.

For the optimum stopping kinematics, the time variable $t'$ is shifted so that the final time is zero.
The initial state for $u$ is $u_s$, and the final state is unknown. 
Therefore, $\alpha=0$ at $t'=0$ holds.
The distance travelled condition is replaced by
\begin{linenomath}
\begin{align}
 \D'' = \lim_{\tau \to \infty} \dfrac{1}{(-\tau)} \int_{-\tau}^0 \f_\text{stop}(t') \dd t' - 1 = 0.
\end{align}
\end{linenomath}
Following the numerical procedure then yields the optimal stopping kinematics.
Figure~\ref{fig:Collapse} shows that the profiles for finite but large $\eps$ converge to the optimal starting and stopping kinematics.
An empirical but convenient fit
\begin{linenomath}
\begin{align}
\label{eqn:empiricalfit}
 \f_\text{start}(t') = \left(1 - e^{-At'} \right)^{1/4}, ~
 \f_\text{stop}(t')  = \left(1 - e^{Bt'^n} \right)^{1/(4n)},
\end{align}
\end{linenomath}
with $A\approx 2.62$, $B \approx 2.06$ and $n\approx 0.36$ approximates the computed starting and stopping dynamics with correct asymptotic behavior and an error with an $L^1$ norm of <1\%.

A composite expression for the optimal kinematics for $\eps \gg 1$ can now be constructed using $\f_\text{start}$ and $\f_\text{stop}$. 
Because $\f_\text{start} \leq 1$, the distance travelled by the plate always lags behind one moving with cruising speed. 
Indeed, $\int_0^\infty (f(t')-1)~\dd t' \approx -0.14$, implying in dimensional terms that by the time steady cruising at a speed $U_c$ is reached, the optimal kinematics lags a distance $0.14 L\times (U_c T/D)$ behind.
Similarly, an additional distance of $0.43 L\times (U_c T/D)$ is lost when stopping.
Thus, the total distance travelled in time $T$ is $U_c T (1 - 0.57/\eps)$. 
Equating this distance to $D$ yields $U_c = (D/T)/(1 - 0.57/\eps)$. 
The corresponding composite expression for $f(t)$ is
\begin{linenomath}
\begin{align} \label{eqn:composite}
 f(t) \approx \dfrac{\f_\text{start}(\eps t) + \f_\text{stop}(\eps (t-1)) - 1 }{1 - ({0.57}/{\eps})}.
\end{align}
\end{linenomath}

\section{Discussion and conclusion}


We have characterized the fastest motion of a flat plate moving parallel to its surface a fixed distance within a work budget.
The salient results are as follows.
The optimum velocity of the plate starts from rest as $\tr{t}^{1/4}$ and comes to a complete stop as $(T-\tr{t})^{1/4}$.
When the distance travelled is large compared to the plate length, the optimum kinematics consists of an optimum start-up, followed by a uniform cruising and an optimum stopping.
In this case, the minimum time for displacement is given by \eqref{eqn:mainresult}.
The spectral condition is used to prove that the computed local minimum is a global one.

These results, in essence, apply to motion of streamlined bodies such as airfoils where the drag arises from skin friction and can be used for open-loop programming of underwater robotics. 
The state-of-the-art approach in controlling underwater actuators uses a parameterization of the hydrodynamic forces\cite{McMillan1995,Yuh2001,Sivcev2018}, thus eliminating salient features of the optimum kinematics governed by the boundary layer growth.
The solution can also be used as a test for more sophisticated computational implementations of fluid mechanical optimization.

The implication of the spectral condition goes beyond the solution of the brachistochrone.
The derivation of the spectral condition is inspired by elements of the upper-bound theory for Navier-Stokes equations to bound properties of turbulence\cite{Hopf1940,Doering1992,Doering1994,Constantin1995,Nicodemus1997}.
The theory seeks upper bounds on properties of turbulent flow, e.g., the dissipation rate, and one approach uses the Lagrange dual of an optimization problem, with the Navier-Stokes equations as constraints, to derive them\cite{Kerswell1999}.
Analogous to the presentation in \S\ref{sec:cfgo}, the spectral constraint naturally follows on the Lagrange multiplier.

In this article, the principles from the upper-bound theory are modified and applied to computational optimization of fluid flow.
Such optimization employs the adjoint formulation for efficient implementation of gradient descent but suffers from the uncertainity of being trapped in a local minimum.
By leveraging the similarity between the adjoint-based gradient descent and the upper-bound theory, we have presented the spectral condition to address this difficulty.
The biggest drawback of applying the spectral condition is that when it fails, it gives no indication of the underlying reason.
The failure could be because the local optimum is not a global one or because of a duality gap.
More work is needed to be able to distinguish between these possibilities.
When the condition succeeds, as it did for the solution presented here, the possibility of the global optimum being different from the one found is elminated.
The spectral condition computationally amounts to an additional (possibly generalized) eigenvalue problem per time-step of the converged local optimum.
Compared to the iterations for gradient descent the computational effort for testing the spectral condition is marginal.

For the spectral condition, the domain $\Omega$ is assumed to not vary with $t$.
This means that for problems with deforming boundaries, e.g., sloshing\cite{Ibrahim2001,Terashima2001}, shape\cite{Mohammadi2004,Brandenburg2009,Mohammadi2010} and kinematic\cite{Kern2006,Gazzola2012,vanRees2015,Maertens2017} optimization problems, the variable domain needs to be mapped to a fixed reference domain \cite[e.g., see][]{Brandenburg2009}.
When this transformation maintains the quadratic nature of the Navier-Stokes equations, the spectral condition remains applicable.
The coordinates used for the flat-plate brachistochrone considered in \S\ref{sec:brachistochrone} illustrate such a transformation.

The spectral condition can also be used in optimization problems constrained by other quadratic partial differential equations, e.g., the Foppl-von Karman equations for deformation of flat plates\cite{JonesMahadevan2015}, the Kortewg-de Vries equation for waves\cite{Dalphin2018}, the Kuramoto-Sivashinsky\cite{Gomes2016} for reacting flows, and partial differential equations used for image processing\cite{Aubert2006}.
From a fundamental perspective, the fluid mechanical brachistochrone is a prototypical example of fluid mechanical optimization, in the same way as the brachistochrone problem is for calculus of variations. 
In this way, the significance of the solution we have presented surpasses these applications.


\section*{Materials and Methods}
The elimination of the highest $x$ derivatives in the Navier-Stokes equations made by the boundary layer approximation causes a loss of regularity in the imposition of the spectral condition.
A term $\sigma \left\langle v_x^2\right\rangle$, with $\sigma \ll 1$, is added numerically to $\W$ and $\hat\W$ to restore this regularity.
We use $\sigma = 10^{-2}$. 
We use a computational domain $-3 \leq x \leq 3$ and $ 0 \leq y \leq 8$.
Smaller values of $\sigma$ and larger domains do not change the results presented here.
\subsection*{Numerical solution and gradient descent}
\eqref{eqn:prandtl} are discretized on a fixed non-uniform grid in $t$, $y$ and $x$, such that grid points are clustered closer to $t=0$ and $1$, $y=0$ and $x = 0$ and $1$. (A number of different clustering schemes were tested to verify the 2-digit accuracy in the numerical results.)
The partial differential equations \eqref{eqn:prandtl} were discretized using first-order upwind finite differences -- the term $(u+f)u_x$ was treated explicitly, while $vu_y$ and $u_{yy}$ were treated implicitly in time. 
The adjoint equations \eqref{eqn:adjoints} were discretized to be the numerical adjoints of the discretization of \eqref{eqn:prandtl}.
A two-level checkpointing scheme\cite{Griewank2000} is used to generate $u$ and $v$ needed to integrate the adjoint variables backwards in $t$.
This procedure ensures that for any discretized $\f(t)$, the numerical solution satisfies the discretized versions of \eqref{eqn:prandtl}, \eqref{eqn:bcs}, \eqref{eqn:adjoints} and \eqref{eqn:abcs}, and the first variation $\delta \LL/\delta\f$ from \eqref{eqn:gradf} can be calculated.
The optimization in $\f$ is achieved using gradient descent by starting from an initial guess $\f^0(t) = 1$, and iteratively updating it as $\f^{n+1}(t) = \f^n(t) + s ({\delta \LL}/{\delta \f})$, for a fixed small number $s \approx 10^{-3}$.
The multiplier $\lambda$ in \eqref{eqn:gradf} is chosen so that $\D=0$ is satisfied by $\f^{n+1}$.

\subsection*{Numerical verification of the spectral constraint}
The solution of \eqref{eqn:spectraloptimization} is the smallest eigenvalue of the generalized eigenvalue problem $\mathcal{A}\psi = \lambda \mathcal{B}\psi$ where
\begin{linenomath}
\begin{subequations}
\label{eqn:spectraloperators}
\begin{align}
\mathcal{A}\psi &\equiv \psi_{yyyy} + \sigma \psi_{xxxx} - \eps \left[ (\hat\alpha_{*y}\psi_y)_y + \frac{(\hat\alpha_{*x}\psi_x)_y + (\hat\alpha_{*x}\psi_y)_x}{2}  \right] \\
\mathcal{B}\psi &\equiv  \psi - \psi_{xx} - \psi_{yy}.
\end{align}
\end{subequations}
\end{linenomath}
The boundary condition $u=0$ as $x\to\pm\infty$ needs closer examination, because this implies $\hat{u} = f \neq 0$ there, which needs to be solved as part of the eigenvalue problem.
Noting that $u=0$ is equivalent to $\hat{u}_x = u_x=0$, the eigenvalue problem is solved with the latter boundary condition.
The value of $f$ can then be read off from the solution.
As before, these operators are programmed as sparse matrices based on the finite-different discretization.
An implicitly restarted Lanczos algorithm for symmetric matrices from the ARPACK library\cite{Lehoucq1998} in shift-invert mode is then used to find the smallest eigenvalue.

\section*{Acknowledgments}

The author is grateful to the Geophysical Fluid Dynamics Summer School at the Woods Hole Oceanographic Institution, where part of this work was performed. Computational resources were provided by the University of Warwick Scientific Computing Research Technology Platform.

%

\end{document}